\documentstyle[12pt]{article}
\makeatletter

\def\section{\@startsection {section}{1}{\z@}{-1.5ex plus -.5ex
minus -.2ex}{1ex plus .2ex}{\large\bf}}

\renewcommand{\thesection}{\arabic{section}.}

\def\@thmcountersep{}

\long\def\@makecaption#1#2{\vskip 10pt \setbox\@tempboxa\hbox{#1. #2}
   \ifdim \wd\@tempboxa >\hsize   
       #1. #2\par                 
     \else                        
       \hbox to\hsize{\hfil\box\@tempboxa\hfil}
   \fi}

\def\ps@headings{
 \def\@oddhead{\footnotesize\rm\hfill\runninghead\hfill}
 \def\@evenhead{\@oddhead}
 \def\@oddfoot{\rm\hfill\thepage\hfill}\def\@evenfoot{\@oddfoot} }

\makeatother

\title{(1+1)-Dimensional Methods for \\General Relativity}{}{}

\def\runninghead{J.H. Yoon :\quad (1+1)-Dimensional Methods}

\author{
{\em Jong-Hyuk Yoon} \thanks{Center for Theoretical Physics and
Department of Physics, Seoul National University, Seoul 151-742,
Korea. e-mail address: SNU00162@KRSNUCC1.Bitnet.
This work was partially supported by the Ministry of Education
and by the Korea Science and Engineering Foundation.}}
\date{} 

\begin{document}
\pagestyle{headings}
\flushbottom

\maketitle
\vspace{-10pt} 

\begin{abstract}

We present the (1+1)-dimensional method for studying
general relativity of 4-dimensions. We first discuss the
general formalism, and subsequently draw attention to the
algebraically special class of space-times, following the
Petrov classification. It is shown that this class of space-times
can be described by the (1+1)-dimensional Yang-Mills action
interacting with matter fields, with the spacial
diffeomorphisms of the 2-surface as the gauge symmetry.
The constraint appears polynomial in part,
whereas the non-polynomial part is a non-linear sigma model type
in (1+1)-dimensions. It is also shown that the representations of
$w_{\infty}$-gravity appear naturally as special cases of
this description, and we discuss briefly the
$w_{\infty}$-geometry in term of the fibre bundle.

\end{abstract}

\renewcommand{\theequation}{\thesection\arabic{equation}}
\setcounter{equation}{0}
\section{Introduction} 

For past years many 2-dimensional field theories have been
intensively studied as laboratories for many theoretical issues,
due to great mathematical simplicities that often exist in
2-dimensional systems.
Recently these 2-dimensional field theories
have received considerable attention, for different reasons,
in connection with general relativistic systems of 4-dimensions,
such as self-dual spaces \cite{par} and the black-hole
space-times \cite{mis,wit}. These 2-dimensional
formulations of self-dual spaces and black-hole space-times of
allow, in principle, many 2-dimensional
field theoretic methods developed in the past relevant
for the description of the physics of 4-dimensions.
This raises an intriguing question as to whether it is also
possible to describe general relativity itself as a
2-dimensional field theory. Recently we
have shown that such a description is indeed possible,
and obtained, at least formally,
the corresponding (1+1)-dimensional action principle
based on the (2+2)-decomposition of general
space-times\footnote[1]{Here we are viewing space-times
of 4-dimensions as locally fibrated, $M_{1+1}\times N_{2}$,
with $M_{1+1}$ as the base manifold of signature $(-,+)$
and $N_{2}$ as the 2-dimensional fibre space of
signature $(+,+)$.}\cite{jhyoon}.
In particular, the algebraically special class of
space-times (the Petrov type II), following the Petrov
classification \cite{pet},
was studied as an illustration from this perspective, and
the (1+1)-dimensional action principle and the constraints for this
class were identified \cite{jhyoona}.
In this (2+2)-decomposition general relativity
shows up as a (1+1)-dimensional gauge theory interacting with
(1+1)-dimensional
matter fields, with the {\it minimal} coupling to the gauge fields,
where the gauge symmetry is the diffeomorphisms of
the fibre of 2-spacial dimensions. In this article we
shall review our recent attempts of the (1+1)-dimensional
formulation of general relativity.
This article is organized as follows. In section 2, we present
the general formalism of (2+2)-decomposition of general relativity,
and establish the corresponding (1+1)-dimensional action principle.

In section 3, we draw attention to the algebraically
special class of space-times \cite{pet,kun}, following the Petrov
classification, and present the (1+1)-dimensional action principle
for this entire class of space-times.
We shall show that the spacial diffeomorphisms of the 2-surface
becomes the gauge fixing condition in this description.
The constraint is polynomial in part,
whereas the non-polynomial term is a non-linear sigma model type
in (1+1)-dimensions. As such, this formulation
might render the problem of the constraints
of general relativity manageable, at least formally.

In section 4, we discuss the realizations of the so-called
$w_{\infty}$-gravity as special cases of this description.
We find the fibre bundle as the natural framework for the
geometric description of $w_{\infty}$-gravity, whose
geometric understanding was lacking so far \cite{wgra,wgrb}.
In this picture the local gauge fields for
$w_{\infty}$-gravity are identified as the connections valued in
the infinite dimensional Lie algebra associated with the
area-preserving diffeomorphisms of the 2-dimensional fibre.
Due to this picture of $w_{\infty}$-geometry,
we are able to construct field theoretic realizations of
$w_{\infty}$-gravity in a straightforward way. In section 5,
we summarize this review and
discuss a few problems for the future investigations.

\renewcommand{\theequation}{\thesection\arabic{equation}}
\setcounter{equation}{0}
\section{$(2+2)$-decomposition of general relativity}  

Consider a 4-dimensional manifold
$P_{4}\simeq M_{1+1} \times N_{2}$, equipped with a metric $g_{A B}$
($A,B, \cdots =0,1,2,3$)\footnote[2]{From here on, we
shall distinguish the two manifolds by their signatures to
avoid confusion. Namely, $M_{1+1}$ shall be referred to as
the (1+1)-dimensional manifold and $N_{2}$ as
2-dimensional manifold.}.
Let $\partial_\mu=\partial/ \partial x ^ \mu$
($\mu, \nu, \cdots = 0,1 $) and
$ \partial_a =\partial/ \partial y ^ a$ $(a, b, \cdots = 2,3)$
be a coordinate basis of $M_{1+1}$ and $N_{2}$, respectively, and
choose $\partial_A=(\partial_\mu, \partial_a)$ as a coordinate
basis of $P_{4}$. In this basis the
most general metric on $P_{4}$ can be written as \cite{c75}
\begin{equation}
ds^2=\phi_{ab}dy^{a}dy^{b}+
     \Big( \gamma_{\mu\nu}+
     \phi_{a b}A_{\mu}^{\ a}A_{\nu}^{\ b}\Big)
     dx^{\mu}dx^{\nu}
     +2\phi_{a b}A_{\mu}^{\ b}dx^{\mu}dy^{a}.     \label{general}
\end{equation}
Formally this is quite similar to the `dimensional reduction' in
Kaluza-Klein theory, where $N_{2}$ is regarded as
the `internal' fibre \footnote[3]{For the algebraically
special class of space-times
we shall consider in section 3, the fibre space $N_{2}$
may be interpreted as
the physical transverse wave-surface \cite{pet}.}
and $M_{1+1}$ as the `space-time'.
In the standard Kaluza-Klein reduction one assumes
a restriction on the metric, namely, an isometry condition,
to make $A_{\mu}^{\ a}$ a gauge field associated
with the isometry group. Here, however, we do not assume
any isometry condition,
and allow all the fields to depend arbitrarily
on both $x^\mu $ and $y^a $.
Nevertheless  $A_{\mu}^{\ a}(x,y)$ can still be
identified as a connection,
but now associated with an infinite dimensional diffeomorphism
group diff$N_{2}$. To show this, let us consider the
following diffeomorphism of $N_{2}$,
\begin{equation}
y^{' a}=y^{' a} (y^{b}, x^{\mu}),
\hspace{1cm}
x^{' \mu}=x^{\mu}.                      \label{new}
\end{equation}
Under these transformations, we find
\renewcommand{\theequation}{\thesection 3\alph{equation}}
\setcounter{equation}{0}
\begin{eqnarray}
& &\gamma '_{\mu\nu}(y' ,x)=\gamma _{\mu\nu}(y,x),   \label{gam}\\
& &\phi'_{a b}(y' ,x)={\partial y^{c} \over \partial y^{' a}}
   {\partial y^{d} \over \partial y^{' b}}
   \phi_{c d}(y,x),                          \label{phi}\\
& &A_{\mu}^{\ 'a}(y',x)={\partial y^{' a} \over \partial y^{c}}
   A_{\mu}^{\ c}(y,x) -\partial_{\mu} y^{' a}.  \label{newtran}
\end{eqnarray}
\renewcommand{\theequation}{\thesection\arabic{equation}}
\setcounter{equation}{3}
\hspace{-.5cm}
For the corresponding infinitesimal variations such that
\begin{equation}
\delta y^{a}=\xi^{a} (y^{b}, x^{\mu}),   \label{var}
\hspace{1cm}
\delta x^{\mu}=0,
\end{equation}
\renewcommand{\theequation}{\thesection 5\alph{equation}}
\setcounter{equation}{0}
\hspace{-.5cm}
(2.3) become
\begin{eqnarray}
\delta \gamma_{\mu\nu}&=&-[ \xi, \gamma_{\mu\nu}]
=-\mbox{\pounds}_{\xi}\gamma_{\mu\nu}
=-\xi^{c}\partial_{c} \gamma_{\mu\nu}, \\
\delta \phi_{a b}&=&-[ \xi, \phi ]_{a b}
=-\mbox{\pounds}_{\xi}\phi_{a b}           \nonumber \\
& =&-\xi^{c}\partial_{c}\phi_{a b}
   -(\partial_{a}\xi^{c})\phi_{c b}
   -(\partial_{b}\xi^{c})\phi_{a c},   \label{var2}\\
\delta A_{\mu}^{\ a}&=&-\partial_{\mu}\xi^{a} + [A_{\mu},\ \xi ]^a
   =-\partial_{\mu}\xi^{a}
    +\mbox{\pounds}_{A_{\mu}}\xi^{a}             \nonumber\\
&=&-\partial_{\mu}\xi^{a}+(A_{\mu}^{c}
    \partial_{c}\xi^{a}
    -\xi^{c}\partial_{c}A_{\mu}^{\ a}), \label{var1}
\end{eqnarray}
\renewcommand{\theequation}{\thesection\arabic{equation}}
\setcounter{equation}{5}
\hspace{-.5cm}
where  $\mbox{\pounds}_{\xi}$ represents the Lie derivative along
the vector fields $\xi=\xi^{a}\partial_a$,
and acts only on the `internal'
indices $a, b,$ etc.  Notice that the
Lie derivative, an {\it infinite}
dimensional generalization of the finite dimensional matrix
commutators, appears naturally.
Clearly (2.4) defines a gauge transformation
which leaves the line element (\ref{general}) invariant.
Associated with this gauge transformation, the {\it covariant}
derivative $D_{\mu}$ is defined by
\begin{equation}
D_{\mu}=\partial_{\mu}-\mbox{\pounds}_{A_{\mu}},     \label{cd}
\end{equation}
where the Lie derivative is taken along the vector field
$A_{\mu}=A_{\mu}^{\ a}\partial_{a}$. With this definition, we have
\begin{equation}
\delta A_{\mu}^{\ a}=-D_{\mu}\xi^{a},
\end{equation}
which clearly indicates that  $A_{\mu}^{\ a}$ is the
gauge field valued in
the infinite dimensional Lie algebra associated with
the diffeomorphisms of $N_{2}$. Moreover
the transformation properties
(\ref{gam}) and (\ref{phi}) show that $\gamma_{\mu\nu}$ and
$\phi_{a b}$ are a scalar and tensor field, respectively, under
diff$N_{2}$.
The field strength $F_{\mu\nu}^{\ \ a}$ corresponding
to $A_{\mu}^{\ a}$ can now be defined as
\begin{equation}
[D_\mu, D_\nu]=-F_{\mu\nu}^{\ \ a}\partial_a=-\{
  \partial_\mu A_{\nu} ^ { \ a}-\partial_\nu
  A_{\mu} ^ { \ a} - [A_{\mu}, A_{\nu}]^a \} \partial_a.\label{fie}
\end{equation}
Notice that the field strength transforms covariantly under the
infinitesimal transformation (\ref{var}),
\begin{equation}
\delta F_{\mu\nu}^{\ \ a}=-[\xi, F_{\mu\nu}]^{a}
  =-\mbox{\pounds}_{\xi}F_{\mu\nu}^{\ \ a}.
\end{equation}
To find the (1+1)-dimensional action principle
of general relativity,
we must compute the scalar curvature of space-times in the
(2+2)-decomposition. For this purpose it is convenient to
introduce the following non-coordinate basis
$\hat{\partial}_{A} = (\hat{\partial}_{\mu} , \hat{\partial}_{a} )$
where \cite{c91}
\begin{equation}
\hat{\partial}_\mu \equiv \partial_\mu
- A_{\mu} ^ {\ a}(x,y)\partial_a,
\hspace{1cm}
\hat{\partial}_a \equiv \partial_a  \ .  \label{cov}
\end{equation}
{}From the definition we have
\begin{equation}
[\hat{\partial}_A, \hat{\partial}_B]=f_{A B} ^ { \
\  \ C}(x,y)\hat{\partial}_C,
\end{equation}
where the structure coefficients $f_{A B} ^ { \  \  \ C}$ are
given by
\begin{eqnarray}
& &f_{\mu\nu} ^ { \  \ a}
   =-F_{\mu\nu} ^ { \  \ a},                  \nonumber\\
& &f_{\mu a} ^ { \  \ b}=-f_{a \mu} ^ { \  \ b}
  =\partial_a A_{\mu} ^ { \ b},                 \nonumber\\
& &f_{A B}^{ \ \ \ C}=0, \hspace{1cm} {\rm otherwise}.\label{str}
\end{eqnarray}
The virtue of this basis is that it brings the metric (\ref{general})
into a block diagonal form
\begin{equation}
g_{A B}=\left(\matrix{\gamma_{\mu\nu}  &
0 \cr 0 & \phi_{ab} \cr }\right),
\end{equation}
which drastically simplifies the computation of the scalar curvature.
In this basis the Levi-Civita connections are given by
\begin{equation}
\Gamma_{A B} ^ { \  \  \ C} = {1\over 2}g ^ {C D}
   (\hat{\partial}_A g_{B D}
   +\hat{\partial}_B g_{A D}
   -\hat{\partial}_D g_{A B})
   +{1\over 2}g ^ {C D}(f_{A B D}-f_{B D A}-f_{A D B}),
\end{equation}
where $f_{A B C}=g_{C D}f_{A B} ^ { \  \  \ D}$.
For completeness, we present the connection coefficients
in components,
\begin{eqnarray}
& &\Gamma_{\mu\nu} ^ { \  \ \alpha}
  ={1\over 2}\gamma ^ {\alpha\beta}\Big( \hat{\partial}_\mu
  \gamma_{\nu\beta}
  +\hat{\partial}_\nu \gamma_{\mu\beta}-\hat{\partial}_\beta
   \gamma_{\mu\nu} \Big),                    \nonumber\\
& &\Gamma_{\mu\nu} ^ { \  \ a}
  =-{1\over 2}\phi ^ {a b}\partial_b \gamma_{\mu\nu}
  -{1\over 2}F_{\mu\nu} ^ { \  \ a},                   \nonumber\\
& &\Gamma_{\mu a} ^ { \  \ \nu}=\Gamma_{a\mu} ^ { \  \ \nu}=
  {1\over 2}\gamma ^ {\nu\alpha}\partial_a \gamma_{\mu\alpha}
  +{1\over 2}\gamma ^ {\nu\alpha}\phi_{a b}
  F_{\mu\alpha} ^ { \  \ b},                  \nonumber\\
& &\Gamma_{\mu a} ^ { \  \ b}={1\over 2}\phi ^ {b c}
   \hat{\partial}_\mu \phi_{a c}
   +{1\over 2}\partial_a A_{\mu} ^ { \ b}-{1\over 2}
   \phi ^ {b c}\phi_{a e}\partial_c A_{\mu} ^ { \ e}, \nonumber\\
& &\Gamma_{a \mu} ^ { \  \ b}={1\over 2}\phi ^ {b c}
   \hat{\partial}_\mu \phi_{a c}
   -{1\over 2}\partial_a A_{\mu} ^ { \ b}-{1\over 2}
    \phi ^ {b c}\phi_{a e}\partial_c A_{\mu} ^ { \ e}, \nonumber\\
& &\Gamma_{a b} ^ { \  \ \mu}
=-{1\over 2}\gamma ^ {\mu\nu}\hat{\partial}_\nu \phi_{a b}
 +{1\over 2}\gamma ^ {\mu\nu}\phi_{a c}\partial_b A_{\nu} ^ { \ c}
 +{1\over 2}\gamma ^ {\mu\nu}\phi_{b c}
 \partial_a A_{\nu} ^ { \ c},                       \nonumber\\
& &\Gamma_{a b} ^ { \  \ c}={1\over 2}\phi ^ {c d}
  \Big( \partial_a \phi_{b d}
  +\partial_b \phi_{a d}-\partial_d \phi_{a b}\Big). \label{final}
\end{eqnarray}
For later purposes it is useful to have the following identities,
\renewcommand{\theequation}{\thesection 16\alph{equation}}
\setcounter{equation}{0}
\begin{eqnarray}
& &\Gamma_{\alpha\mu}^{\ \ \alpha}
  ={1\over 2}\gamma ^ {\alpha\beta}\hat{\partial}_\mu
  \gamma_{\alpha\beta},
\hspace{1cm}
\Gamma_{a \mu}^{\ \ a}={1\over 2}\phi ^ {a b}
  \hat{\partial}_\mu \phi_{a b}
  -\partial_a A_{\mu} ^ { \ a},           \label{contra1}    \\
& &\Gamma_{\beta a}^{\ \ \beta}={1\over 2}\gamma^{\alpha\beta}
  \partial_a \gamma_{\alpha\beta},
\hspace{1cm}
 \Gamma_{b a}^{\ \ b}={1\over 2}\phi^{b c}
 \partial_a \phi_{b c}.                \label{contra2}
\end{eqnarray}
\renewcommand{\theequation}{\thesection\arabic{equation}}
\setcounter{equation}{16}
\hspace{-.5cm}
The curvature tensors are defined as
\begin{eqnarray}
& &R_{A B C} ^ { \  \  \  \  \ D} =
   \hat{\partial}_{A}^{\phantom{}} \Gamma_{B C} ^ { \  \  \ D}
   -\hat{\partial}_{B}^{\phantom{}} \Gamma_{A C} ^ { \  \  \ D}
   +\Gamma_{A E} ^ { \  \  \ D}\Gamma_{B C} ^ { \  \  \ E}
   -\Gamma_{B E} ^ { \  \  \ D}\Gamma_{A C} ^ { \  \  \ E}
   -f_{A B} ^ { \  \  \ E}\Gamma_{E C} ^ { \  \  \ D}, \nonumber\\
& & R_{A C}=R_{A B C}^{ \  \  \  \  \ B},
\hspace{1cm}
R=g^{A C}R_{A C}.                \label{rie}
\end{eqnarray}
Explicitly, the scalar curvature $R$ is given by
\begin{equation}
R  = \gamma ^ {\mu\nu}( R_{\mu\alpha\nu} ^ { \ \ \ \ \alpha}
 +R_{\mu a\nu} ^ { \ \ \ \ a})
 +\phi ^ {a b}(R_{a c b} ^ { \  \  \ c}
 + R_{a\mu b} ^ { \  \  \ \mu}) \ , \label{sca}
\end{equation}
which becomes, after a lengthy computation,
\begin{eqnarray}
R&=&\gamma ^ {\mu\nu}R_{\mu\nu}+\phi ^ {a c} R_{a c}
   +{ 1\over 4}\phi_{a b}\gamma^{\mu\nu}\gamma^{\alpha\beta}
   F_{\mu\alpha} ^ { \  \ a}F_{\nu\beta}^{\ \  b}   \nonumber\\
& &   +{1\over 4}\gamma ^ {\mu\nu}\phi ^ {a b}\phi ^ {c d}
     \Big\{
     (D_{\mu}\phi_{a c})(D_{\nu}\phi_{b d})
  -(D_{\mu}\phi_{a b})(D_{\nu}\phi_{c d})\Big\}     \nonumber\\
& & +{1\over 4}\phi ^ {a b}\gamma ^ {\mu\nu}\gamma ^ {\alpha\beta}
   \Big\{
   (\partial_a \gamma_{\mu\alpha})(\partial_b \gamma_{\nu\beta})
    -(\partial_a \gamma_{\mu\nu})
     (\partial_b \gamma_{\alpha\beta})\Big\}
   +\nabla_A j ^ A,                            \label{sca1}
\end{eqnarray}
where $R_{\mu\nu}$ and $R_{a c}$ are defined by
\renewcommand{\theequation}{\thesection 20\alph{equation}}
\setcounter{equation}{0}
\begin{eqnarray}
& &R_{\mu\nu} = \hat{\partial}_{\mu}^{\phantom{}}
    \Gamma_{\alpha\nu} ^ { \  \ \alpha}
   -\hat{\partial}_{\alpha}^{\phantom{}}
    \Gamma_{\mu\nu} ^ { \  \ \alpha}
   +\Gamma_{\mu\beta} ^ { \  \ \alpha}
    \Gamma_{\alpha\nu} ^ { \  \ \beta}
   -\Gamma_{\beta\alpha} ^ { \  \ \beta}
   \Gamma_{\mu\nu} ^ { \  \ \alpha},          \label{tensor} \\
& &R_{a c} = \partial_{a}^{\phantom{}} \Gamma_{b c} ^ { \  \ b}
    -\partial_{b}^{\phantom{}} \Gamma_{a c} ^ { \  \ b}
    +\Gamma_{a d} ^ { \  \ b}\Gamma_{b c} ^ { \  \ d}
    -\Gamma_{d b} ^ { \  \ d}\Gamma_{a c} ^ { \  \ b}. \label{til}
\end{eqnarray}
\renewcommand{\theequation}{\thesection\arabic{equation}}
\setcounter{equation}{20}
\hspace{-.5cm}
The last term in (\ref{sca1}) is given by
\renewcommand{\theequation}{\thesection 21\alph{equation}}
\setcounter{equation}{0}
\begin{eqnarray}
& &\nabla_A j ^ A
   =\nabla_\mu j ^ {\mu}+\nabla_a j ^ a,          \label{sura}\\
& &\nabla_\mu j ^ {\mu}=\Big(
    \hat{\partial}_\mu +\Gamma_{\alpha\mu} ^ { \  \ \alpha}
    +\Gamma_{c\mu} ^ { \  \ c}\Big) j^{\mu},  \label{surb}\\
& &\nabla_a j ^ a=\Big(
    \partial_a +\Gamma_{c a} ^ { \  \ c}
    +\Gamma_{\alpha a} ^ { \  \ \alpha}\Big) j^{a},   \label{sur}
\end{eqnarray}
\renewcommand{\theequation}{\thesection\arabic{equation}}
\setcounter{equation}{21}
\hspace{-.5cm}
where $j^{\mu}$ and $j^{a}$ are given by
\begin{equation}
j^{\mu}=\gamma^{\mu\nu}\Big( \phi ^ {a b}
     \hat{\partial}_\nu \phi_{a b}
    -2\partial_a A_\nu ^ { \ a} \Big),
\hspace{1cm}
j^{a}=\phi ^ {a b}\gamma ^ {\mu\nu}\partial_b
    \gamma_{\mu\nu}.
\end{equation}
That $\nabla_A j ^ A$ is a surface term in the action integral
can be seen easily, using (2.16). For instance
let us show that $\sqrt{-\gamma}\sqrt{\phi}\nabla_\mu j ^ {\mu}$
is a surface term, where $\gamma ={\rm det}\gamma_{\mu\nu}$ and
$\phi ={\rm det}\phi_{a b}$. From (\ref{surb}) we have
\begin{eqnarray}
\sqrt{-\gamma}\sqrt{\phi}\nabla_\mu j ^ {\mu}
  =\sqrt{-\gamma}\sqrt{\phi}
  \Big[
   \partial_{\mu} j^{\mu}- A_{\mu}^{\ a}\partial_{a}j^{\mu}
   + \Big( \Gamma_{\alpha\mu} ^ { \  \ \alpha}
   +\Gamma_{c\mu} ^ { \  \ c} \Big) j^{\mu} \Big]. \label{surfa}
\end{eqnarray}
The first term in the r.h.s. of (\ref{surfa}) can be written as
\begin{equation}
\sqrt{-\gamma}\sqrt{\phi}\partial_{\mu}j^{\mu}
  =-{1\over 2}\sqrt{-\gamma}\sqrt{\phi}\Big(
    \gamma^{\alpha\beta}
    \partial_{\mu}\gamma_{\alpha\beta}
   + \phi^{a b}\partial_{\mu}\phi_{a b} \Big) j^{\mu}
   +\partial_{\mu}\Big(
    \sqrt{-\gamma}\sqrt{\phi}j^{\mu}\Big), \label{surfb}
\end{equation}
and for the second term, we have
\begin{equation}
\sqrt{-\gamma}\sqrt{\phi}A_{\mu}^{\ a}\partial_{a}j^{\mu}
=-\sqrt{-\gamma}\sqrt{\phi} \Big[ \Big\{
  A_{\mu}^{\ a}(
  \Gamma_{\alpha a}^{\ \ \alpha} + \Gamma_{b a}^{\ \ b})
  + \partial_{a}A_{\mu}^{\ a} \Big\} j^{\mu} \Big]
 +\partial_{a}
  \Big(
  \sqrt{-\gamma}\sqrt{\phi}A_{\mu}^{\ a}j^{\mu}\Big). \label{surfc}
\end{equation}
The last two terms in the r. h. s. of (\ref{surfa})
becomes, using (2.16),
\begin{eqnarray}
\sqrt{-\gamma}\sqrt{\phi}\Big( \Gamma_{\alpha\mu} ^ { \  \ \alpha}
   +\Gamma_{c\mu} ^ { \  \ c} \Big) j^{\mu}
&=&\sqrt{-\gamma}\sqrt{\phi}\Big(
 {1\over 2}
  \gamma^{\alpha\beta} \partial_{\mu}\gamma_{\alpha\beta}
 -A_{\mu}^{\ a}\Gamma_{\alpha a}^{\ \ \alpha}
 +{1\over 2}
  \phi^{a b}\partial_{\mu}\phi_{a b}         \nonumber\\
& & -A_{\mu}^{\ a}\Gamma_{b a}^{\ \ b}
 -\partial_{a}A_{\mu}^{\ a} \Big) j^{\mu}.        \label{detail}
\end{eqnarray}
Putting (\ref{surfb}), (\ref{surfc}), and (\ref{detail})
into (\ref{surfa}), we find that it is a total divergence term,
\begin{equation}
\sqrt{-\gamma}\sqrt{\phi}\nabla_\mu j ^ {\mu}
 =\partial_{\mu}\Big( \sqrt{-\gamma}\sqrt{\phi}j^{\mu} \Big)
 -\partial_{a}\Big( \sqrt{-\gamma}\sqrt{\phi}
 A_{\mu}^{\ a}j^{a} \Big),       \label{diver}
\end{equation}
which we may ignore. Similarly,
$\sqrt{-\gamma}\sqrt{\phi}\nabla_a j ^ a$
is also a surface term. This altogether shows that
$\sqrt{-\gamma}\sqrt{\phi}\nabla_A j ^ A$ is
indeed a total divergence term.

At this point it is important to notice the followings.
First, $D_{\mu}\phi_{a b}$, written as
\begin{eqnarray}
D_{\mu}\phi_{a b} &= & \partial_\mu \phi_{a b}
  -\mbox{\pounds}_{A_{\mu}}\phi_{a b}           \nonumber\\
&=& \partial_\mu \phi_{a b}
   - \Big\{A_\mu ^ { \ c}(\partial_c \phi_{a b})
   +(\partial_a A_\mu ^ { \ c})\phi_{c b}
   +(\partial_b A_\mu ^ { \ c})\phi_{a c}\Big\} \
\end{eqnarray}
indeed transforms covariantly under the infinitesimal
diffeomorphism (\ref{var}),
\begin{equation}
\delta (D_{\mu}\phi_{a b})=-\mbox{\pounds}_{\xi}(D_{\mu}\phi_{a b})
        =-[\xi, D_{\mu}\phi]_{a b}.
\end{equation}
Second, the derivative $\hat{\partial}_{\mu}$,
when applied to $\gamma_{\mu\nu}$,
becomes the covariant derivative
\begin{equation}
\hat{\partial}_{\mu} \gamma_{\alpha\beta}
=\partial_{\mu}\gamma_{\alpha\beta}
-\mbox{\pounds}_{A_{\mu}}\gamma_{\alpha\beta}
=D_{\mu}\gamma_{\alpha\beta},                  \label{hat}
\end{equation}
so that $\hat{\partial}_{\mu} \gamma_{\alpha\beta}$
transforms covariantly
\begin{equation}
\delta (\hat{\partial}_{\mu} \gamma_{\alpha\beta})
=-\mbox{\pounds}_{\xi}(D_{\mu}\gamma_{\alpha \beta})
=-[\xi, D_{\mu}\gamma_{\alpha \beta}].
\end{equation}
These observations play an important role when we discuss the
gauge invariance of the theory under diff$N_{2}$.
It is worth mentioning here that,
from (\ref{tensor}) and (\ref{hat}),
$R_{\mu\nu}$ becomes the `covariantized' Ricci tensor
\begin{equation}
R_{\mu\nu} = D_\mu
    \Gamma_{\alpha\nu}^{\  \ \alpha}
   -D_\alpha \Gamma_{\mu\nu}^{\ \ \alpha}
   +\Gamma_{\mu\beta}^{ \ \ \alpha}\Gamma_{\alpha\nu}^{\ \ \beta}
   -\Gamma_{\beta\alpha}^{ \ \ \beta}\Gamma_{\mu\nu}^{\ \ \alpha},
\end{equation}
as $\Gamma_{\mu\nu}^{\ \ \alpha}$'s do not involve the
`internal' indices $a,b$, etc. Thus we might call
$\gamma^{\mu\nu}R_{\mu\nu}$ as the
`gauged' gravity action in (1+1)-dimensions \cite{spin}.

With the scalar curvature at hand, one can easily
write down the lagrangian for the Einstein-Hilbert action on $P_{4}$.
{}From (\ref{sca1}) we have
\begin{eqnarray}
{\cal L}_{2}&=&-\sqrt{-\gamma}\sqrt{\phi} \Big[
       \gamma ^ {\mu\nu}R_{\mu\nu}+\phi ^ {a b} R_{a b}
       +{1\over 4}\phi_{a b}F_{\mu\nu} ^ { \  \ a}
       F ^ {\mu\nu b}                        \nonumber\\
& &+{1\over 4}\gamma ^ {\mu\nu}\phi ^ {a b}\phi ^ {c d}
      \Big\{
      (D_{\mu}\phi_{a c})(D_{\nu}\phi_{b d})
     -(D_{\mu}\phi_{a b})(D_{\nu}\phi_{c d})\Big\} \nonumber\\
& &+{1\over 4}\phi ^ {a b}\gamma ^ {\mu\nu}\gamma ^ {\alpha\beta}
  \Big\{
  (\partial_a \gamma_{\mu\alpha})(\partial_b \gamma_{\nu\beta})
  -(\partial_a \gamma_{\mu\nu})(\partial_b \gamma_{\alpha\beta})
  \Big\} \Big],    \label{sca2}
\end{eqnarray}
neglecting the total divergence term ({\ref{diver}).
Clearly the action principle
describes a (1+1)-dimensional field theory which is invariant under
the gauge transformation of diff$N_{2}$,
as the gauge field  $A_{\mu}^{\ a}$
couples {\it minimally} to both $\gamma_{\mu\nu}$ and
$\phi_{a b}$. Therefore each term in (\ref{sca2})
is invariant under diff$N_{2}$.
To understand the physical contents of the theory
we notice the followings.
First, unlike the ordinary gravity, the
metric $\gamma_{\mu\nu}$ of $M_{1+1}$ here is `charged',
because it couples to $A_{\mu}^{\ a}$ (with
the coupling constant 1).
Second, the metric $\phi_{a b}$ of $N_{2}$ can be
identified as a non-linear sigma field, whose self-interaction
potential is determined by the scalar
curvature $\phi^{a b}R_{a b}$ of $N_{2}$.
The theory therefore describes a gauge theory of
diff$N_{2}$ interacting
with the `gauged' gravity and the non-linear sigma field
on $M_{1+1}$.

\renewcommand{\theequation}{\thesection\arabic{equation}}
\setcounter{equation}{0}
\section{Algebraically special class of space-times} 

In contrast to the cases of the self-dual spaces and
black-hole space-times, the (1+1)-dimensional action principle
for general space-times, as we derived in the previous section,
appears to be rather formal and consequently,
of little practical use.
In this section we therefore
draw attention following the Petrov classification
to a specific class of space-times, namely, the algebraically
special class, and interpret the entire class
from the (1+1)-dimensional point of view. It turns out that
space-times of this class can be formulated
as (1+1)-dimensional field theory in a remarkably simple form.

Let us consider a class of space-times that contain a
twist-free null vector field $k^{A}$.
These space-times belong to
the algebraically special class of space-times, according
to the Petrov classification. This class of
space-times is rather broad, since most of the known
exact solutions of the Einstein's equations
are algebraically special.
Being twist-free, the null vector field may be chosen to be a
gradient field, so that $k_{A}=\partial_{A} u$ for
some function $u$. The null hypersurface $N_{2}$
defined by $u={\rm constant}$
spans the 2-dimensional subspace for which we introduce
two space-like coordinates $y^{a}$. The
general line element for this class has the form \cite{pet,kun}
\begin{equation}
ds^{2}=\phi_{a b}dy^{a}dy^{b}
       -2du(dv + m_{a}dy^{a} + H du), \label{metr}
\end{equation}
where $v$ is the affine parameter, and $\phi_{a b}$, $m_{a}$ and
$H$ are functions of all of the four coordinates ($u, v, y^{a}$),
as we assume no Killing vector fields.

For the class of space-times (\ref{metr}),
we shall find the (1+1)-dimensional action principle defined on
the $(u, v)$-surface. For this purpose let us first
introduce the `light-cone' coordinates $(u, v)$ such that
\begin{equation}
u={1\over \sqrt{2}}(x^{0} + x^{1}), \hspace{1cm}
v={1\over \sqrt{2}}(x^{0} - x^{1}),              \label{nul}
\end{equation}
and define $A_{u}^{\ a}$ and $A_{v}^{\ a}$
\begin{equation}
A_{u}^{\ a}={1\over \sqrt{2}}(A_{0}^{\ a} + A_{1}^{\ a}),
\hspace{1cm}
A_{v}^{\ a}={1\over \sqrt{2}}
(A_{0}^{\ a} - A_{1}^{\ a}).                     \label{nula}
\end{equation}
For $\gamma_{\mu\nu}$, we assume the Polyakov ansatz \cite{pol}
\begin{equation}
\gamma_{\mu\nu}=\left( \matrix{ -2h & -1 \cr -1 & 0 \cr }
         \right),
\hspace{1cm}
\gamma^{\mu\nu}=\left( \matrix{ 0 & -1 \cr -1 & 2h \cr }
         \right),
\hspace{1cm}
({\rm det}\gamma_{\mu\nu}=-1),               \label{pol}
\end{equation}
in the $(u, v)$-coordinates.
Then the line element (\ref{general}) becomes
\begin{eqnarray}
ds^{2}& = & \phi_{a b}dy^{a}dy^{b} - 2 du dv -2h (du)^{2}
+ \phi_{a b} (A_{u}^{\ a} du +A_{v}^{\ a} dv)
(A_{u}^{\ b} du +A_{v}^{\ b} dv)    \nonumber\\
 & & + 2 \phi_{a b}(A_{u}^{\ a} du
     + A_{v}^{\ a} dv)dy^{b}.                  \label{res}
\end{eqnarray}
If we choose the `light-cone' gauge \footnote[4]{
Here we are referring to the disposable gauge degrees of
freedom in the action. There could be topological obstruction
against globalizing this choice, as the general coordinate
transformation of $N_{2}$ corresponds
to the gauge transformation.}
$A_{v}^{\ a}=0$, then this becomes
\begin{equation}
ds^{2}=\phi_{a b}dy^{a}dy^{b} - 2\ du \Big[
dv -  \phi_{a b}A_{u}^{\ b} dy^{a}
+\Big( h -{1\over 2}\phi_{a b}A_{u}^{\ a}A_{u}^{\ b}  \Big)du
\Big].                               \label{resa}
\end{equation}
A comparison of (\ref{metr}) and (\ref{resa}) tells us that if the
following identifications
\begin{equation}
m_{a}=-\phi_{a b}A_{u}^{\ b},
\hspace{1cm}
H=h-{1\over 2}\phi_{a b}A_{u}^{\ a}A_{u}^{\ b}  \label{com}
\end{equation}
are made, then the two line elements are the same.
This shows that the Polyakov ansatz (\ref{pol})
amounts to the restriction
(modulo the gauge choice $A_{v}^{\ a}=0$)
to the algebraically special class of space-times
that contain a twist-free null vector field.

Let us now examine the transformation properties of
$h$, $\phi_{a b}$, $A_{u}^{\ a}$, and $A_{v}^{\ a}$
under the diffeomorphism of $N_{2}$,
\begin{equation}
y^{' a}=y^{' a} (y^{b}, u, v),
\hspace{1cm}
u'=u,
\hspace{1cm}
v'=v.                               \label{coor}
\end{equation}
Under these transformations, we find that
\renewcommand{\theequation}{\thesection 9\alph{equation}}
\setcounter{equation}{0}
\begin{eqnarray}
& &h'(y' , u, v)=h(y, u, v), \\
& &\phi'_{a b}(y', u, v)={\partial y^{c} \over \partial y^{' a}}
   {\partial y^{d} \over \partial y^{' b}}\phi_{c d}(y, u, v),   \\
& &A_{u}^{\ 'a}(y', u, v)={\partial y^{' a} \over \partial y^{c}}
   A_{u}^{\ c}(y, u, v) -\partial_{u} y^{' a},  \\
& &A_{v}^{\ 'a}(y' , u, v)=-\partial_{v} y^{' a},  \label{tran}
\end{eqnarray}
\renewcommand{\theequation}{\thesection\arabic{equation}}
\setcounter{equation}{9}
\hspace{-.5cm}
which become, under the infinitesimal variations,
$\delta y^{a}=\xi^{a} (y, u, v)$ and
$\delta x^{\mu}=0$,
\renewcommand{\theequation}{\thesection 10\alph{equation}}
\setcounter{equation}{0}
\begin{eqnarray}
& &\delta h=-[\xi,  h] = -\xi^{a}\partial_{a} h, \\
& &\delta \phi_{a b}=-[\xi,  \phi]_{a b}
   =-\xi^{c}\partial_{c}\phi_{a b}
   -(\partial_{a}\xi^{c})\phi_{c b}
   -(\partial_{b}\xi^{c})\phi_{a c},         \\
& &\delta A_{u}^{\ a}=-D_{u}\xi^{a}
   =-\partial_{u}\xi^{a} + [A_{u}, \xi]^{a},      \label{au} \\
& &\delta A_{v}^{\ a}=-\partial_{v}\xi^{a}.           \label{inf}
\end{eqnarray}
\renewcommand{\theequation}{\thesection\arabic{equation}}
\setcounter{equation}{10}
\hspace{-.5cm}
This shows that $h$ and $\phi_{a b}$ are a scalar and
tensor field, respectively, and
$A_{u}^{\ a}$ and $A_{v}^{\ a}$ are the gauge fields
valued in the infinite dimensional Lie algebra associated
with the group of diffeomorphisms of $N_{2}$.
That $A_{v}^{\ a}$ is a pure gauge is clear, as
it depends on the gauge function $\xi^{a}$ only. Therefore
it can be always set to zero, at least locally,
by a suitable coordinate transformation (\ref{coor}).
To maintain the explicit gauge
invariance, however, we shall work with the line element
(\ref{res}) in the following,
with the understanding that $A_{v}^{\ a}$ is a pure gauge.

Let us now proceed to write down the action
principle for (\ref{res})
in terms of the fields $h$, $\phi_{a b}$, $A_{u}^{\ a}$, and
$A_{v}^{\ a}$. For this purpose, it is convenient to decompose
the 2-dimensional metric $\phi_{a b}$ into the conformal
classes
\begin{equation}
\phi_{a b}=\Omega\rho_{a b},
\hspace{1cm}
(\Omega > 0 \  \  {\rm and} \ \
{\rm det}\rho_{a b} = 1).                    \label{dec}
\end{equation}
The kinetic term $K$ of $\phi_{a b}$ in (\ref{sca2}) then becomes
\begin{eqnarray}
K&\equiv &{1\over 4}\sqrt{-\gamma}\sqrt{\phi}
   \gamma ^ {\mu\nu}\phi ^ {a b}\phi ^ {c d}
         \Big\{
        (D_{\mu}\phi_{a c})(D_{\nu}\phi_{b d})
        -(D_{\mu}\phi_{a b})(D_{\nu}\phi_{c d})\Big\}   \nonumber\\
&=&-{ (D_{\mu} \Omega)^{2}\over 2\Omega }
  + {1\over 4}\Omega\gamma^{\mu\nu}\rho^{a b}\rho^{c d}
    (D_{\mu}\rho_{a c})(D_{\nu}\rho_{b d})      \nonumber\\
&=&-{1\over 2}{\rm e}^{\sigma}(D_{\mu}\sigma)^{2}
   + {1\over 4}{\rm e}^{\sigma}\gamma^{\mu\nu}\rho^{a b}\rho^{c d}
    (D_{\mu}\rho_{a c})(D_{\nu}\rho_{b d}),       \label{kin}
\end{eqnarray}
where we defined $\sigma$ by $\sigma={\rm ln}\Omega$, and
the covariant derivatives $D_{\mu}\Omega$, $D_{\mu}\rho_{a b}$,
and $D_{\mu}\sigma$ are
\renewcommand{\theequation}{\thesection 13\alph{equation}}
\setcounter{equation}{0}
\begin{eqnarray}
& &D_{\mu}\Omega = \partial_{\mu}\Omega
  - A_{\mu}^{\ a}\partial_{a}\Omega
  - (\partial_{a}A_{\mu}^{\ a}) \Omega,         \\
& &D_{\mu}\rho_{a b}=\partial_{\mu}\rho_{a b}
  - [A_{\mu}, \rho]_{a b}
  + (\partial_{c}A_{\mu}^{\ c})\rho_{a b},         \label{rho}\\
& &D_{\mu}\sigma=\partial_{\mu}\sigma
  -  A_{\mu}^{\ a}\partial_{a}\sigma
  - \partial_{a}A_{\mu}^{\ a},                     \label{den}
\end{eqnarray}
\renewcommand{\theequation}{\thesection\arabic{equation}}
\setcounter{equation}{13}
\hspace{-.5cm}
respectively,
where $[A_{\mu}, \rho]_{a b}$ is given by
\begin{equation}
[{A_{\mu}}, \rho]_{ a b}=
    A_\mu ^ { \ c}\partial_c \rho_{a b}
    +(\partial_a A_\mu ^ {\ c})\rho_{c b}
    +(\partial_b A_\mu ^ { \ c})\rho_{a c}.       \label{rhoa}
\end{equation}
The inclusion of the divergence term
$\partial_{a}A_{\mu}^{\ a}$ in (3.13) is necessary to
ensure (3.13) transform covariantly (as the tensor fields)
under diff$N_{2}$, since $\Omega$ and $\rho_{a b}$ are the tensor
densities of weight $-1$ and $+1$, respectively.
Using the ansatz (\ref{pol}), the kinetic term
(\ref{kin}) becomes
\begin{eqnarray}
K&=&{\rm e}^{\sigma}(D_{+}\sigma)(D_{-}\sigma)
   - {1\over 2}{\rm e}^{\sigma}\rho^{a b}\rho^{c d}
    (D_{+}\rho_{a c})(D_{-}\rho_{b d})   \nonumber\\
& &-h{\rm e}^{\sigma}\Big\{ (D_{-}\sigma)^{2}
   - {1\over 2}\rho^{a b}\rho^{c d}
    (D_{-}\rho_{a c})(D_{-}\rho_{b d})  \Big\},     \label{s}
\end{eqnarray}
where $+ (-)$ stands for $u (v)$. The Polyakov ansatz
(\ref{pol}) simplifies enormously
the remaining terms in the action (\ref{sca2}),
as we now show.
Let us first notice that ${\rm det}\gamma_{\mu\nu}=-1$.
Therefore the term
\begin{equation}
\sqrt{-\gamma}\sqrt{\phi}\phi^{a c}R_{a c}
=\sqrt{\phi}\phi^{a c}R_{a c}
\end{equation}
can be removed from the action being a surface term.
Moreover, since we have
\begin{equation}
\gamma^{\mu\nu}\partial_{a}\gamma_{\mu\nu}
={2\over \sqrt{-\gamma}}\partial_{a} \sqrt{-\gamma}=0,
\end{equation}
the last term in the action (\ref{sca2}) vanishes.
Furthermore, one can easily verify that
\begin{eqnarray}
\phi^{a b}\gamma^{\mu\nu}\gamma^{\alpha\beta}
   (\partial_a \gamma_{\mu\alpha})
   (\partial_b \gamma_{\nu\beta})
&=&\phi^{a b}(\partial_a \gamma_{++})
   \gamma^{+-}(\partial_b \gamma_{- \alpha})
   \gamma^{\alpha +}                      \nonumber\\
&=&0,
\end{eqnarray}
since $\partial_{b}\gamma_{- \alpha}=0$.
The only remaining terms that
contribute to the
action (\ref{sca2}) are thus
the (1+1)-dimensional Yang-Mills action and the `gauged'
gravity action. The Yang-Mills action becomes
\begin{equation}
{1\over 4}\phi_{a b}F_{\mu\nu}^{\ \ a}F^{\mu\nu b}
=-{1\over 2}{\rm e}^{\sigma}\rho_{a b}
F_{+ -}^{\ \ a}F_{+ -}^{\ \ b}.              \label{yang}
\end{equation}
To express the `gauged' Ricci scalar $\gamma^{\mu\nu}R_{\mu\nu}$
in terms of $h$ and $A_{v}^{\ a}$, etc., we have to compute the
Levi-Civita connections first. They are given by
\begin{eqnarray}
& &\Gamma_{++}^{\ \ +}=-D_{-}h,
\hspace{1cm}
\Gamma_{++}^{\ \ -}=D_{+}h + 2h D_{-}h, \nonumber\\
& &\Gamma_{+-}^{\ \ -}=\Gamma_{-+}^{\ \ -}=D_{-}h,  \label{levia}
\end{eqnarray}
and vanishing otherwise. Thus the `gauged' Ricci tensor becomes
\begin{equation}
R_{+-}=R_{-+}=-D_{-}^{2} h,
\hspace{1cm}
R_{--}=0.                       \label{rica}
\end{equation}
{}From (\ref{pol}) and (\ref{rica}), the `gauged'
Ricci scalar $\gamma^{\mu\nu} R_{\mu\nu}$  is given by
\begin{equation}
\gamma^{\mu\nu} R_{\mu\nu}=2\gamma^{+-} R_{+-}
=2D_{-}^{2} h,                                \label{scal}
\end{equation}
since $\gamma^{++}=R_{--}=0$. Putting together (\ref{s}),
(\ref{yang}), and (\ref{scal}) into (\ref{sca2}), the
action becomes
\begin{eqnarray}
{\cal L}_{2}& = &
-{1\over 2}{\rm e}^{2\sigma}\rho_{a b}F_{+-}^{\ \ a}F_{+-}^{\ \ b}
+{\rm e}^{\sigma}(D_{+}\sigma) (D_{-}\sigma)
-{1\over 2}{\rm e}^{\sigma}\rho^{a b}\rho^{c d}
(D_{+}\rho_{a c})(D_{-}\rho_{b d})     \nonumber\\
& &+h{\rm e}^{\sigma}\Big\{
{1\over 2}\rho^{a b}\rho^{c d}
(D_{-}\rho_{a c})(D_{-}\rho_{b d}) -(D_{-}\sigma)^{2} \Big\}
+2{\rm e}^{\sigma} D_{-}^{2} h.            \label{mast}
\end{eqnarray}
The last term in (\ref{mast}) can be expressed as
\begin{eqnarray}
{\rm e}^{\sigma} D_{-}^{2} h
&=&{\rm e}^{\sigma}
   \Big( \partial_{-} - A_{-}^{\ b}\partial_{b} \Big)
   \Big( \partial_{-} h
   -  A_{-}^{\ a}\partial_{a} h \Big) \nonumber\\
&=&{\rm e}^{\sigma}
   \Big\{ \partial_{-}^{2} h -
   \partial_{-}\Big( A_{-}^{\ a}\partial_{a} h \Big)
   -A_{-}^{\ a}\partial_{a}(D_{-}h) \Big\}         \nonumber\\
&=&-(\partial_{-}{\rm e}^{\sigma})(\partial_{-} h)
   +(\partial_{-}{\rm e}^{\sigma})
   \Big( A_{-}^{\ a}\partial_{a} h \Big)
   +\partial_{a}\Big(
   {\rm e}^{\sigma}A_{-}^{\ a} \Big) (D_{-}h) \nonumber\\
& & +\partial_{-}\Big( {\rm e}^{\sigma}\partial_{-} h \Big)
 -\partial_{-}\Big( {\rm e}^{\sigma}A_{-}^{\ a}\partial_{a} h\Big)
 -\partial_{a}\Big( {\rm e}^{\sigma}A_{-}^{\ a}D_{-}h
   \Big)                                   \nonumber\\
&\simeq &-{\rm e}^{\sigma} (\partial_{-} \sigma)(D_{-} h)
  +{\rm e}^{\sigma}A_{-}^{\ a}(\partial_{a} \sigma)(D_{-}h)
  +{\rm e}^{\sigma}(\partial_{a}A_{-}^{\ a})(D_{-} h) \nonumber\\
&=&- {\rm e}^{\sigma} (D_{-} \sigma)(D_{-} h),    \label{sto}
\end{eqnarray}
where we dropped the surface term and used (\ref{den}).
This can be written as
\begin{eqnarray}
{\rm e}^{\sigma} (D_{-} \sigma)(D_{-} h)
&=&{\rm e}^{\sigma} (D_{-} \sigma)
   \Big( \partial_{-} h - A_{-}^{\ a}\partial_{a} h \Big)\nonumber\\
&=&-h\partial_{-}\Big(  {\rm e}^{\sigma}D_{-}\sigma   \Big)
   +h\partial_{a}\Big(
   {\rm e}^{\sigma}A_{-}^{\ a} D_{-}\sigma \Big)
   +\partial_{-}
   \Big( h {\rm e}^{\sigma} D_{-}\sigma  \Big)     \nonumber\\
& &-\partial_{a} \Big( h {\rm e}^{\sigma}
    A_{-}^{\ a} D_{-}\sigma \Big)                \nonumber\\
&\simeq & -h {\rm e}^{\sigma} ( \partial_{-}\sigma )(D_{-}\sigma)
   - h {\rm e}^{\sigma} \partial_{-} ( D_{-} \sigma )
   + h {\rm e}^{\sigma}A_{-}^{\ a}(\partial_{a} \sigma)
     (D_{-}\sigma)                \nonumber\\
& &+ h {\rm e}^{\sigma} (\partial_{a}A_{-}^{\ a})(D_{-}\sigma)
   + h {\rm e}^{\sigma}A_{-}^{\ a}
     \partial_{a}(D_{-}\sigma)                   \nonumber\\
&=&-h{\rm e}^{\sigma} \Big\{  D_{-}^{2}\sigma
   + (D_{-}\sigma)^{2}   \Big\}.       \label{addition}
\end{eqnarray}
We therefore have
\begin{equation}
{\rm e}^{\sigma} D_{-}^{2}\ h \simeq h {\rm e}^{\sigma} \Big\{
D_{-}^{2}\sigma + (D_{-}\sigma)^{2}   \Big\},    \label{surface}
\end{equation}
neglecting the surface terms.
The resulting (1+1)-dimensional action principle therefore becomes
\begin{eqnarray}
{\cal L}_{2}& = &
-{1\over 2}{\rm e}^{2\sigma}\rho_{a b}F_{+-}^{\ \ a}F_{+-}^{\ \ b}
+{\rm e}^{\sigma}(D_{+}\sigma) (D_{-}\sigma)
-{1\over 2}{\rm e}^{\sigma}\rho^{a b}\rho^{c d}
(D_{+}\rho_{a c})(D_{-}\rho_{b d})           \nonumber\\
& &+h{\rm e}^{\sigma}\Big\{
2D_{-}^{2}\sigma + (D_{-}\sigma)^{2}
+ {1\over 2}\rho^{a b}\rho^{c d}
(D_{-}\rho_{a c})(D_{-}\rho_{b d}) \Big\},           \label{two}
\end{eqnarray}
up to the surface terms.
Notice that $h$ is a Lagrange multiplier, whose variation
yields the constraint
\begin{equation}
H_{0}=D_{-}^{2}\sigma + {1\over 2}(D_{-}\sigma)^{2}
 + {1\over 4}\rho^{a b}\rho^{c d} (D_{-}\rho_{a c})(D_{-}\rho_{b d})
\approx 0.                                    \label{ham}
\end{equation}
{}From this (1+1)-dimensional point of view, $h$ is the
lapse function (or a pure gauge) that prescribes how to `move
forward in the $u$-time', carrying the surface $N_{2}$
at each point of the section $u={\rm constant}$.
The constraint, $H_{0}\approx 0$, is
{\it polynomial} in $\sigma$ and $A_{-}^{\ a}$,
and contains a non-polynomial term of the non-linear sigma model
type but in (1+1)-dimensions, where such models often admit
exact solutions. This allows us to view
the problem of the constraints of general
relativity \cite{con} from a new perspective.

We now have the (1+1)-dimensional action principle
for the algebraically special class of space-times that contain
a twist-free null vector field. It is described by the
Yang-Mills action, interacting with the fields
$\sigma$ and $\rho_{a b}$ on the `flat' (1+1)-dimensional surface,
which however must satisfy the
constraint $H_{0} \approx 0$.
(The flatness of the (1+1)-dimensional surface can be seen
from the fact that the lapse function, $h$, can be chosen
as zero, provided that $H_{0} \approx 0$ holds.)
The infinite dimensional group of the diffeomorphisms of
$N_{2}$ is {\it built-in} as the local gauge symmetry,
via the minimal couplings to the gauge fields.

Having formulated the algebraically special class of space-times
as a gauge theory on (1+1)-dimensions, we may wish to apply
varieties of field theoretic methods developed in (1+1)-dimensions.
For instance, the action (\ref{two}) can be viewed as the
bosonized form \cite{bos} of {\it some} version of
the (1+1)-dimensional QCD in the infinite dimensional
limit of the gauge group \cite{mig}. For small fluctuations
of $\sigma$, the action (\ref{two}) becomes
\begin{equation}
{\cal L}_{2} =
 -{1\over 2}\rho_{a b}F_{+-}^{\ \ a}F_{+-}^{\ \ b}
 +(D_{+}\sigma)(D_{-}\sigma) -{1\over 2}\rho^{a b}\rho^{c d}
 (D_{+}\rho_{a c})(D_{-}\rho_{b d}),               \label{small}
\end{equation}
modulo the constraint $H_{0}\approx 0$. It is beyond the
scope of this article to investigate these theories as
(1+1)-dimensional quantum field theories. However, this
formulation raises many intriguing questions such as: would there
be any phase transition in quantum gravity as viewed as the
(1+1)-dimensional quantum field theories? If it does, then what does
that mean in quantum geometrical terms?
Thus, general relativity, as viewed from the (1+1)-dimensional
perspective, renders itself to be studied as a gauge theory
in full sense \cite{gau}, at least for the class of space-times
discussed here.

\renewcommand{\theequation}{\thesection\arabic{equation}}
\setcounter{equation}{0}
\section{$w_{\infty}$-gravity as special cases} 

In the previous section we derived the action principle
on (1+1)-dimensions as the vantage point of studying
general relativity for this algebraically special
class of space-times.
We now ask different but related questions: what kinds of
other (1+1)-dimensional field theories related to this problem
can we study? For these, let us consider
the case where the local gauge symmetry is replaced by the
area-preserving diffeomorphisms
of $N_{2}$. (For these varieties of field theories,
we shall drop the
constraint (\ref{ham}) for the moment. It is at this point
that we are departing from general relativity.)
This class of field theories naturally realizes the so-called
$w_{\infty}$-gravity \cite{wgra,wgrb} in a
linear and geometric way, as we now describe.

The area-preserving diffeomorphisms are generated
by the vector fields
$\xi^{a}$, tangent to the surface $N_{2}$ and divergence-free,
\begin{equation}
\partial_{a} \xi^{a}=0.                      \label{div}
\end{equation}
Let us find the gauge fields $A_{\pm}^{\ a}$ compatible with
the divergence-free condition (\ref{div}).
Taking the divergence of both sides of (\ref{au})
and (\ref{inf}), we have
\begin{equation}
\partial_{a}\delta A_{\pm}^{\ a}
=-\partial_{\pm}(\partial_{a}\xi^{a})
+\partial_{a}[ A_{\pm}, \xi ]^{a}.      \label{coma}
\end{equation}
This shows that the condition $\partial_{a}A_{\pm}^{\ a}=0$
is invariant under the area-preserving diffeomorphisms,
and characterizes a special subclass of the gauge fields,
compatible with the condition (\ref{div}).
Moreover, when $\partial_{a}A_{\pm}^{\ a}=0$, the fields
$\rho_{a b}$ and $\sigma$ behave under the
area-preserving diffeomorphisms
as a tensor and a scalar field, respectively,
as (\ref{rho}) and (\ref{den}) suggest. Indeed,
the Jacobian for the area-preserving
diffeomorphisms is just 1, disregarding the distinction
between the tensor fields and the tensor densities.
The (1+1)-dimensional action principle now becomes
\begin{equation}
{\cal L}_{2}' =
-{1\over 2}{\rm e}^{2\sigma}\rho_{a b}F_{+-}^{\ \ a}F_{+-}^{\ \ b}
+{\rm e}^{\sigma}(D_{+}\sigma) (D_{-}\sigma)
-{1\over 2}{\rm e}^{\sigma}\rho^{a b}\rho^{c d}
(D_{+}\rho_{a c})(D_{-}\rho_{b d}),              \label{area}
\end{equation}
where $D_{\mu}\sigma$, $D_{\mu}\rho_{a b}$, and $F_{+-}^{\ \ a}$ are
\renewcommand{\theequation}{\thesection 4\alph{equation}}
\setcounter{equation}{0}
\begin{eqnarray}
& &D_{\pm}\sigma=\partial_{\pm}\sigma -  A_{\pm}^{\ a}
   \partial_{a}\sigma,               \label{z}    \\
& &D_{\pm}\rho_{a b}=\partial_{\pm} \rho_{a b}
  - [A_{\pm}, \rho]_{a b},                  \\
& &F_{+-}^{\ \ a}=\partial_{+} A_{-} ^ {\ a}
  -\partial_{-} A_{+} ^ { \ a} - [A_{+}, A_{-}]^{a}.  \label{areb}
\end{eqnarray}
\renewcommand{\theequation}{\thesection\arabic{equation}}
\setcounter{equation}{4}
\hspace{-.5cm}
Under the infinitesimal variations
\begin{equation}
\delta y^{a}=\xi^{a}(y, u, v),
\hspace{1cm}
\delta x^{\mu}=0,
\hspace{1cm}
(\partial_{a}\xi^{a}=0),
\end{equation}
the fields transform as
\renewcommand{\theequation}{\thesection 6\alph{equation}}
\setcounter{equation}{0}
\begin{eqnarray}
& &\delta \sigma=-[\xi, \sigma]=-\xi^{a}\partial_{a}\sigma,    \\
& &\delta \rho_{a b}=-[\xi, \rho]_{a b}
   =-\xi^{c}\partial_{c}\rho_{a b}
   -(\partial_{a}\xi^{c})\rho_{c b}-
    (\partial_{b}\xi^{c})\rho_{a c}, \\
& &\delta A_{+}^{\ a}=-D_{+}\xi^{a}
   =-\partial_{+}\xi^{a} + [A_{+}, \xi]^{a},          \\
& &\delta A_{-}^{\ a}=-\partial_{-}\xi^{a},
\end{eqnarray}
\renewcommand{\theequation}{\thesection\arabic{equation}}
\setcounter{equation}{6}
\hspace{-.45cm}
which shows that it
{\it is} a linear realization of the
area-preserving diffeomorphisms. The geometric picture of
the action principle (\ref{area})
is now clear: it is equipped with the natural bundle structure,
where the gauge fields are the connections valued in the
Lie algebra associated with the area-preserving
diffeomorphisms of $N_{2}$.
Thus the action principle (\ref{area}) provides a
field theoretical realization of
$w_{\infty}$-gravity \cite{wgra,wgrb}
in a linear and geometric way, with the built-in area-preserving
diffeomorphisms as the local gauge symmetry.

With this picture of $w_{\infty}$-geometry at hands, we may
construct as many different realizations of $w_{\infty}$-gravity
as one wishes. The simplest example would be a single real
scalar field representation, which we may write
\begin{equation}
{\cal L}_{2}'' = -{1\over 2}F_{+-}^{\ \ a}F_{+-}^{\ \ a}
+(D_{+}\sigma) (D_{-}\sigma),                     \label{sing}
\end{equation}
where we used $\delta_{a b}$ in the summation, and
$D_{\pm}\sigma$ and $F_{+-}^{\ \ a}$ are as given in
(\ref{z}) and (\ref{areb}).
By choosing the gauge $A_{-}^{\ a}=0$ and eliminating the auxiliary
field $A_{+}^{\ a}$ in terms of $\sigma$ using the equations of
motion of $A_{+}^{\ a}$, we recognize (\ref{sing}) a single
real scalar field realization of $w_{\infty}$-gravity.
In presence of the auxiliary field $A_{+}^{\ a}$,
(\ref{sing}) provides an example of the
{\it linearized} realization of $w_{\infty}$-gravity
for a single real scalar field.
It would be interesting to see if the representation (\ref{sing})
is related to the ones constructed in
the literatures \cite{wgra,wgrb}.

\section{Discussion} 

In this review, we examined space-times of 4-dimensions
from a (1+1)-dimensional point of view. That general relativity
admits such a description is rather surprising, even though
the action principle in general appears rather formal. For the
algebraically special class of space-times, however, the
(1+1)-dimensional action principle, as we have shown here,
is formulated as the Yang-Mills type gauge theories
interacting with matter fields,
where the infinite dimensional group of diffeomorphisms of
the 2-surface becomes the `internal' gauge symmetry.
The  constraint conjugate to the lapse function
appears partly as polynomial.
The non-polynomial part is a typical non-linear sigma model type in
(1+1)-dimensions, where such models often admit exact
solutions. We also discussed the so-called
$w_{\infty}$-gravity as special cases of the algebraically
special class of space-times.
The detailed study of the $w_{\infty}$-gravity and its geometry
in terms of the fibre bundle will be presented
somewhere else.

We wish to conclude with a few remarks. First, one might
be interested in finding exact solutions of the Einstein's equations
in this formulation. Various two (or more) Killing reductions of
the Einstein's equations have been known
for sometime which led to the
discovery of many exact solutions to the Einstein's equations,
by making the system essentially two (or lower) dimensional.
In our formulation, the Einstein's equations are already put
into a two dimensional form without such assumptions.
This might be useful in finding new solutions of the
Einstein's equations, which possess no Killing
symmetries\footnote[5]{Interestingly,
there {\it are} exact solutions of the Einstein's equations which
possess no {\it space-time} Killing symmetry, known as the
Szekeres' dust solutions \cite{pet}. For the vacuum Einstein's
equations, however, no such solutions are known, at least to
the author.}.

Second, we need to find the constraint algebras for the
algebraically special class of
space-times explicitly in terms of the variables we used here.
As we have shown here, the splitting of
the metric variables into the
gauge fields and the `matter fields' is indeed
suitable for the description of general relativity
as Yang-Mills type gauge theories in (1+1)-dimensions.
It remains to study the constraint algebras
in detail to see if the ordering problem in
the constraints of general relativity becomes manageable
in terms of these variables.

Lastly, that the Lie algebra of $SU(N)$ for large $N$
can be used as an approximation of the infinite dimensional
Lie algebra of the area-preserving
diffeomorphisms of the 2-surface has been suggested as a way of
`regulating' the area-preserving diffeomorphisms.
In connection with the problem regarding the regularization
of quantum gravity in this formulation,
one might wonder as to whether it is also
possible to approximate the
diffeomorphism algebras of the 2-surface in terms of finite
dimensional Lie algebras in a certain limit. There seem to be
many interesting questions to be asked about general relativity
in this formulation.

\bibliographystyle{plain}

\end{document}